\documentclass[12pt,a4paper]{article}
\usepackage{jheppub}

\usepackage{graphicx}
\usepackage{subcaption}
\usepackage{xspace}

\usepackage[utf8]{inputenc}

\usepackage[style=numeric-comp,sorting=none]{biblatex}
\abstract{At the heart of experimental high energy physics (HEP) is the development of facilities and instrumentation that provide sensitivity to new phenomena. Our understanding of nature at its most fundamental level is advanced through the analysis and interpretation of data from sophisticated detectors in HEP experiments.  The goal of data analysis systems is to realize the maximum possible scientific potential of the data within the constraints of computing and human resources in the least time. To achieve this goal, future analysis systems should empower physicists to access the data with a high level of interactivity, reproducibility and throughput capability. As part of the HEP Software Foundation’s Community White Paper process, a working group on Data Analysis and Interpretation was formed to assess the challenges and opportunities in HEP data analysis and develop a roadmap for activities in this area over the next decade. In this report, the key findings and recommendations of the Data Analysis and Interpretation Working Group are presented.}

\begin{document}

\noindent
\begin{tabular*}{\linewidth}{lc@{\extracolsep{\fill}}r@{\extracolsep{0pt}}}
 & & HSF-CWP-2017-05 \\
 & & March 25, 2018 \\ 
 & & \\
\end{tabular*}
\vspace{1.0cm}

\title{HEP Software Foundation Community White Paper Working Group -- Data Analysis and Interpretation}

\author{HEP Software Foundation:}
\author[a]{Lothar Bauerdick}
\author[b]{Riccardo Maria Bianchi}
\author[c]{\\ Brian Bockelman}
\author[d,e]{Nuno Castro}
\author[f]{Kyle Cranmer}
\author[g]{Peter Elmer}
\author[h]{\\ Robert Gardner}
\author[i]{Maria Girone}
\author[a,1]{Oliver Gutsche}
\author[i]{Benedikt Hegner}
\author[j]{\\ Jos\'{e} M. Hern\'{a}ndez}
\author[a]{Bodhitha Jayatilaka}
\author[g]{David Lange}
\author[k,1]{Mark S. Neubauer}
\author[k]{Daniel S. Katz}
\author[l]{Lukasz Kreczko}
\author[m]{James Letts}
\author[n]{Shawn McKee}
\author[o]{\\ Christoph Paus}
\author[a]{Kevin Pedro}
\author[g]{Jim Pivarski}
\author[p]{Martin Ritter}
\author[q]{Eduardo Rodrigues}
\author[l]{Tai Sakuma}
\author[a]{Elizabeth Sexton-Kennedy}
\author[p]{Michael D. Sokoloff}
\author[r]{\\ Carl Vuosalo}
\author[m]{Frank W\"{u}rthwein}
\author[s]{Gordon Watts}

\affiliation[a]{Fermi National Accelerator Laboratory, Batavia, Illinois USA}
\affiliation[b]{University of Pittsburgh, Pittsburgh, Pennsylvania, USA}
\affiliation[c]{University of Nebraska-Lincoln, Lincoln, Nebraska, USA}
\affiliation[d]{Laboratório de Instrumentação e Física Experimental de Partículas, Lisbon, Portugal}
\affiliation[e]{University of Minho, Braga, Portugal}
\affiliation[f]{New York University, New York, New York, USA}
\affiliation[g]{Princeton University, Princeton, New Jersey, USA}
\affiliation[h]{University of Chicago, Chicago, Illinois, USA}
\affiliation[i]{CERN, Geneva, Switzerland}
\affiliation[j]{Centro de Investigaciones Energéticas, Medioambientales y Tecnológicas, Madrid, Spain}
\affiliation[k]{University of Illinois at Urbana-Champaign, Urbana, Illinois USA}
\affiliation[l]{University of Bristol, Bristol, UK}
\affiliation[m]{University of California at San Diego, La Jolla, California, USA}
\affiliation[n]{University of Michigan, Ann Arbor, Michigan, USA}
\affiliation[o]{Massachusetts Institute of Technology, Cambridge, Massachusetts, USA}
\affiliation[p]{Fakultät für Physik, Ludwig-Maximilians-Universität München, München, Germany}
\affiliation[q]{University of Cincinnati, Cincinnati, Ohio, USA}
\affiliation[r]{University of Wisconsin-Madison, Madison, Wisconsin, USA}
\affiliation[s]{University of Washington, Seattle, Washington, USA}
\affiliation[1]{Paper Editor}

\maketitle

\newpage

\section{Introduction}

High energy physics (HEP) answers scientific questions by analyzing the data obtained from detectors and sensors of suitably designed experiments, comparing measurements with predictions from models and theories. The number of different questions that can be answered by a single experiment range from a few for scientifically focused devices to a very diverse and large set of questions for multi-purpose devices. In all cases, data are analyzed by groups of researchers of varying sizes, from individual researchers to very large groups of scientists. ``Analysis'' as described here includes the reduction of processed detector data, from the trigger and event reconstruction algorithms, to create, analyze and interpret estimates of specific physics observables. The baseline analysis model utilizes successive stages of data reduction, finally analyzing a compact dataset with quick real-time iteration. Experiments use a series of processing steps to reduce large input datasets down to sizes suitable for laptop-scale analysis. The line between managed production-like analysis processing and individual analysis, as well as the balance between common versus individualized analysis data formats differs by experiment based on their needs, degree of optimization and the maturity of an experiment in its life cycle. The current baseline model stems from the goal to exploit the maximum possible scientific potential of the data while minimizing the ``time-to-insight'' for a large number of different analyses performed in parallel. It is a complicated product of diverse criteria ranging from computing resources and related innovation to management styles of the experiment collaborations. An evolution of the baseline approach is the ability to produce physics-ready data right from the output of the high-level trigger of the experiment, whereas the baseline approach also depends on further processing of the data with updated or new software algorithms or detector conditions. This could be a key enabler of a simplified analysis model that allows very efficient data reduction.

The goal of data analysis systems is to realize the maximum possible scientific potential of the data within the constraints of computing and human resources in the least time.
Analysis models aim towards giving scientists access to the data in the most interactive way possible, and to enable quick turn-around in iteratively learning new insights from the data. Many analyses have common deadlines defined by conference schedules. The increased analysis activity before these deadlines requires the analysis system to be sufficiently elastic to guarantee a rich set of physics results. Also, heterogeneous computing hardware like GPUs, FPGAs, and new memory architectures are emerging and can be exploited to further reduce the ``time-to-insight''.

As part of the HEP Software Foundation’s Community White Paper~\cite{HSF-CWP-2017-01} process, a working group on Data Analysis and Interpretation was formed to assess the challenges and opportunities in HEP data analysis and develop a roadmap for activities in this area over the next decade. In this report, the key findings and recommendations of the Data Analysis and Interpretation Working Group are presented.

\section{HEP Analysis Software Ecosystem}

Over the past 20 years, the HEP community has developed and gravitated around a single analysis ecosystem: ROOT~\cite{Brun1996}. This ecosystem currently dominates HEP analysis and impacts the full event processing chain, providing key elements such as foundation libraries and I/O services. It has been empowering for the HEP community as ROOT provides an integrated and validated toolkit which lowers the bar for analysis, enables the community to speak in a common analysis language and facilitates the generation of improvements and additions to the toolkit that are rapidly propogated to the whole community to benefit a large number of analyses. On the other hand, the landscape of open source software (both academic and industrial) for data analysis is rapidly evolving and surpasses HEP in terms of development effort and community size, compelling a community re-evaluation of this strategy.

The emergence and abundance of alternative and new analysis components and techniques coming from industry open source projects is a challenge for the HEP analysis software ecosystem. The HEP community is very interested in using these new techniques and technologies, especially together with established components of the ecosystem where possible. It should also be possible to interchange old components with new open source components. We propose in the first year to perform R\&D on enabling new open source tools to be dynamically plugged into the existing ecosystem and mechanisms to freely exchange parts of the ecosystem with new components. This could include investigating new ways of package management and distribution following open source approaches. On a 3-year time frame, we propose to research a comprehensive set of {\it bridges} and {\it ferries} between the HEP analysis ecosystem and the industry analysis tool landscape, where a bridge enables the ecosystem to use an open source analysis tool and a ferry allows to use data from the ecosystem in the tool and vice versa.

The maintenance and sustainability of the current analysis ecosystem is a challenge. The ecosystem supports a number of use cases and both integrates and maintains a wide variety of components. Components have to be prioritized to fit into the available effort envelope, which is provided by a few institutions and less distributed across the community. Legacy and less used parts of the ecosystem are hard to retire and their continued support strain the available effort. In the first year, we propose R\&D to evolve policies to minimize this effort by retiring less used components from the integration and validation efforts. We propose to enable individuals to continue to use retired components by taking over their maintenance and validation following the central efforts of the ecosystem, spending a little of their own effort. 
But not every component can just be retired if it is not used anymore by most of the ecosystem users. Therefore for the 3-year time frame, we propose to evolve our policies regarding adoption of alternatives in the analysis community and how to retire old components of the ecosystem.

\section{Analysis Languages}

Given the large prior investment in software and the predominance of ROOT-based data analysis models in HEP experiments, when considering reconstruction and other performance-critical code, the community will continue to look to C++. Recent developments within the data science community, however, lead us to additional considerations regarding software languages in support of data analysis.

\subsection{Python}
For analysis code, Python has developed into the language of choice in the data science community and also increasingly within the HEP community. It is widely used in the data analytics community in other fields of science and industry and is liked for fast development cycles and ease-of-use. Python also has a rich ecosystem of well-maintained and advanced software packages. Python should become a first-class language in HEP, but performance-critical code should be offloaded to libraries, written in the most suitable language for the problem domain, thus allowing optimization ``under the hood'' by experts. Popular examples of such libraries are Numpy~\cite{5725236-numpy}, pandas~\cite{mckinney-proc-scipy-2010}, and Tensorflow~\cite{abadi2016tensorflow}.  Experiences from inside and outside the HEP community indicate that the simpler interfaces and code constructs of Python could reduce the complexity of analysis code and therefore contribute to the decrease in ``time-to-insight'' for HEP analyses. For example, data frames in pandas can represent arbitrary dimensions of histograms and offer a rich set of operations on the data, such as the split-apply-combine strategy~\cite{JSSv040i01}. The community needs to ensure that scientists at all levels are included in the community language choice and not left behind, and to do so, education and documentation need to be a strong component of making Python a first-class language in HEP. We propose for the first year time frame to finalize full support of Python in our ecosystem including long term maintenance.

\subsection{Declarative Languages} 
Further optimization could be gained by employing a functional or declarative programming model. This would allow scientists to express the intended data transformation as a query on data. Instead of an analyzer defining and controlling the ``how'' of the analysis workflow, the analyst would declare the ``what'', essentially removing the need to define the event loop in an analysis and leave it to underlying services and systems to optimally iterate over events. In analogy to how programming in C++ abstracts implementation features compared to programming in assembler, it appears that these high-level approaches will allow one to abstract from the underlying implementations, providing computing systems more freedom to optimize use of the underlying resources. We propose to assess the advantages of the usage of a declarative programming model for data analysis over cognitive overhead for newcomers. More concretely, we propose to provide to the analysis community demonstrators to express their analysis following a declarative approach. Such demonstrators should not only help to assess the viability of the programming model but also attempt to fully leverage hardware resources at disposal, heterogeneous or not, both locally and in a distributed fashion. We propose on the 3-year time frame to conclude on the already ongoing R\&D projects (e.g. TDataFrame in ROOT) and to follow up with additional R\&D projects to develop a prototype functional or declarative programming language model. 

\section{Analyzing Data}

\subsection{File format and compression algorithms} 
The baseline approach to analyzing HEP data roughly follows a common pattern. The data are prepared for analysis by first completing most of the CPU-intensive processing steps and then making the resulting data available to the community. In subsequent analysis steps, the I/O performance for iterating over the data becomes one of the driving factors in minimizing the ``time-to-insight'' for analyses. There are many file format standards used by a wealth of data analytics tools in other science fields and industry which should be explored for use in HEP analyses, but it is generally felt that currently ROOT I/O is a good fit to the community's needs. Disk space is usually the key concern of the experiment computing models as disk is the most expensive hardware component. The community extensively uses compression to minimize storage costs. This reduces the performance of purely iterating over the data due to the cost of decompression. To improve on the current state and to enable analysis of much larger datasets than today, R\&D will be needed to continue in file formats, compression algorithms, and new ways of storing and accessing data for analysis. 

\subsection{Analysis Facilities} 
Towards the era of the High-Luminosity LHC (HL-LHC)~\cite{Apollinari:2284929}, we envision dedicated ``smart'' facilities optimized for data analysis, which provide an extendable environment that in turn provides fully functional analysis capabilities. Such end-to-end engineered facilities will give users easy and well-maintained access to the HEP analysis software ecosystem and to experiment-specific capabilities (software, documentation, databases, {\it etc}). They will support complex ``from beginning to end'' analysis workflows including support for provenance tracking, continuous integration, verification, and reproducibility. They will provide low-latency, high-throughput computing capabilities with access to complete data sets without the user having to fall back to the high-latency ``batch system style'' analysis. ``Primitive'' versions of such analysis facilities are currently provided e.g. at CERN, Fermilab and other places. However for the HL-LHC, such dedicated analysis facilities would provide a much more end-to-end engineered latency-optimized and maintained environment, that then also would provide focus points for support and training of analysts.

We propose to focus technology developments to support analysis by prototyping such analysis facilities to develop the capabilities and the required ecosystem to support the analysis use cases mentioned in Section~\ref{sec:models}. This prototyping work  includes to continue to investigate optimizing the storage systems and used representation of data on disk together, and also to facilitate the utilization of new additional storage layers like SSD storage and NVRAM-like storage, which exhibit different characteristics than the currently dominant spinning disk installations. This should include a fresh look at the concept of ``virtual data'', optimizing the choice between stored versus re-computed data products. Regarding a work plan for designing and deploying data analysis facilities to support HL-LHC data analysis, much of the required developments can be efficiently driven by a small number of use cases, that, over the course of 5 years, should be refined and realigned with the help of demonstrators and analysis challenges, using an iterative approach towards prototyping such a facility (or facilities, for the different experiments). Detailed plans should be worked out regarding the prototyping of the capabilities of analysis facilities, iteratively scaling up both functionalities and capacities, up to the capabilities required for the HL-LHC. In Years 1 to 3, the conceptualization of an end-to-end low-latency response high-capacity analysis facility should be started, with an initial prototype. In Years 3 to 5, analysis of data from the 3rd running period of the LHC should be used to evaluate chosen architectures and verify design or provide input for corrective actions. At the end of this process starting in Year 5, a blueprint for remaining developments and system design should become available, in time for deployment.
\subsection{Non-event data handling} 
Another area that has not yet received the attention it deserves is the access to non-event data for analysis (cross section values, scale factors, tagging efficiencies, etc.). The community feels that the existing capabilities for event data, namely easy storage of event data of all sorts of different content, a similar way of saving and accessing non-event information during the analysis step is needed. There exist many ways of doing this now, but no commonly accepted and supported way has yet emerged. This could be expanded to consider event versus non-event data in general to support use cases from small data volumes (for example cross sections) to large data volumes (BDTs and NNs). We propose R\&D in the area of non-event information handling on the 3-year time scale, which would facilitate analysis at much higher scales than today.

\section{Analysis Models and Future Systems}\label{sec:models}

Methods for analyzing the data at the LHC experiments have been developed over the years and successfully applied to LHC data to produce physics results during Run-1 and Run-2. Analysis at the LHC experiments typically starts with users running code over centrally-managed data that is of O(100 kB/event) and contains all of information required to perform a typical analysis leading to publication. In this section, we describe some of the proposed analysis models for the future, building on the experiences of the past.

\subsection{Sequential Ntuple Reduction}

The most common approach to analyzing data is through a campaign of data reduction and refinement, ultimately producing data frames (flat ntuples) and histograms used to make plots and tables from which physics inference can be made. In the following, we are using ATLAS and CMS computing model to illustrate data volumes involved in LHC analyses. The centrally-managed data are O(100 kB/event) and are typically too large (e.g. O(100 TB) for 35 fb$^{-1}$ of Run-2 data) to be analyzed by users on their local computing infrastructures like their workgroup servers or laptops. An often stated aim of the data reduction steps is to arrive at a dataset that ``can fit on one's laptop'', presumably to facilitate low-latency, high-rate access to a manageable amount of data during the final stages of analysis.  At its core, creating and retaining intermediate datasets from a data reduction campaign, bringing and keeping them ``close'' (e.g. on laptop/desktop) to the analyzers, is designed to minimize latencies and risks related to resource contention.

\subsection{``Spark''-like Analysis}

A new model of data analysis, developed outside of HEP, maintains the concept of sequential ntuple reduction but mixes interactivity with batch processing. Spark is one such system, but TensorFlow~\cite{abadi2016tensorflow}, Dask~\cite{rocklin2015dask}, Blaze~\cite{wiebe2014blaze}, Parsl~\cite{babuji_yadu_2017_853492}, Pachyderm~\cite{thepachydermteam}, and Thrill~\cite{bingmann2016thrill} are others. Distributed processing is either launched as a part of user interaction at a command prompt or wrapped up for batch submission. The key differences from the above are:
\begin{itemize}
\item Parallelization is implicit through map/filter/reduce functionals or more general workflow patterns and constructs,
\item Data are abstracted as remote, distributed datasets, rather than files,
\item Computation and storage are mixed for data locality: a specialized cluster must be prepared, but can yield higher throughput.
\end{itemize}

A Spark-like analysis facility would be a shared resource for exploratory data analysis (e.g. making quick plots on data subsets through the spark-shell) and batch submission with the same interface (e.g. substantial jobs through spark-submit). The primary advantage that software products like Spark introduce is in simplifying the user's access to data, lowering the cognitive overhead of setting up and running parallel jobs. Certain types of jobs may also be faster than batch processing, especially flat ntuple processing (which benefits from SQL-like optimization) and iterative procedures such as fits and machine learning (which benefit from cluster-wide cache).

Although Spark itself is the leading contender for this type of analysis, as it has a well-developed ecosystem with many third-party tools developed by industry, it is the style of analysis workflow that we are distinguishing here rather than the specific technology present today. Spark itself is hard to interface with C++, but this might be alleviated by projects such as ROOT's TDataFrame, which presents a Spark-like interface in ROOT, and may allow for more streamlined interoperability.

\subsection{Query-based Analysis}

An alternative to the previously described models is to perform fast querying of  centrally-managed data and compute remotely on the queried data to produce the analysis products of interest, especially but not limited to histograms. This is similar to NoSQL databases tuned for exploratory data analysis, such as Spark-SQL~\cite{Armbrust:2015:SSR:2723372.2742797}, Impala~\cite{bittorf2015impala}, Kudu~\cite{t.lipcond.alvesd.burkertj.d.cryansa.dembom.percys.rusd.wangm.bertozzic.p.mccabea.wang.2015}, Hawq~\cite{Chang:2014:HMP:2588555.2595636}, Apache Drill~\cite{drill}, and Google Dremel/BigQuery~\cite{36632}, but with a focus on distribution plotting, rather than SQL table transformation.

\begin{figure}
\begin{center}
\includegraphics[width=\textwidth]{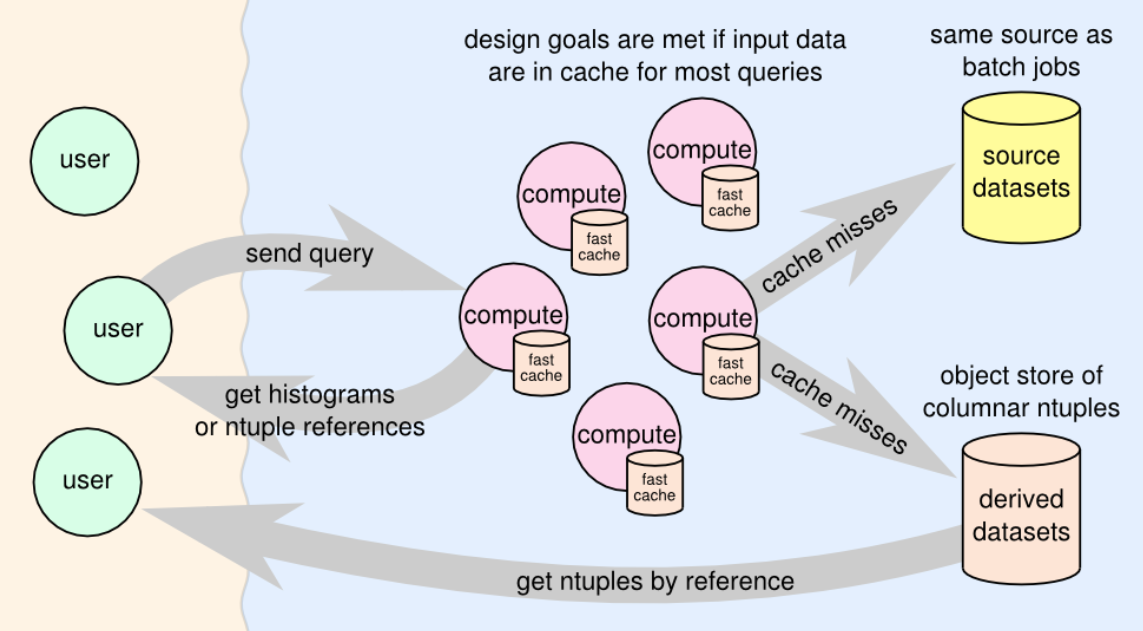}
\caption{\label{fig:query_system}Schematic depiction of a possible future query-based analysis system.}
\end{center}
\end{figure}

In one vision for a query-based analysis approach, a series of analysis cycles, each of which provides minimal input (queries of data and code to execute), generates the essential output (histograms, ntuples, etc.) which can be retrieved by the user. The analysis workflow should be accomplished without focus on persistence of data traditionally associated with data reduction. However, transient data may be generated in order to efficiently accomplish this workflow and optionally could be retained to facilitate an analysis ``checkpoint'' for subsequent execution. In this approach, the focus is on obtaining the analysis end-products in a way that does not necessitate a data reduction campaign and associated provisioning of resources. A schematic depiction of a possible future query-based analysis system is shown in Fig~\ref{fig:query_system}.

\vskip 0.1in\noindent
Advantages of a query-based analysis include:

\vskip 0.1in\noindent
\textbf{\textit{Minimalistic Analysis}}.
A critical consideration of the Sequential Ntuple Reduction method might reasonably question why analyzers would bother to generate and store intermediate data to get to the same outcomes of interest (histograms, etc). A more economical approach is to provide only the minimal information -- code providing instructions for selecting the dataset, events of interest, and items to plot. 

\vskip 0.1in\noindent
\textbf{\textit{Democratization of Analyses}}.
In the Sequential Ntuple Reduction method, as one gets further down the data reduction chain, the user (or small group of users) needs to figure out how to provision and manage the storage required to accommodate this intermediate data which in many cases is accessed with small ($<10^{-4}$) or zero duty cycle. For small groups, the resources required (both in personnel and hardware) to execute such a data reduction campaign might be prohibitive in the HL-LHC era, effectively ``pricing them out'' of contributing strongly to analyses -- possibly a lost opportunity for innovation and discovery. Removing the requirements on storing intermediate data in the analysis chain would help to ``democratize'' data analysis and streamline the overall analysis workflow.

\vskip 0.1in\noindent
\textbf{\textit{Ease of Provenance}}.
The query-based analysis provides an opportunity for autonomous storage of provenance information, as all processing in an analysis step from ``primary'' analysis-level data to the histograms is contained to a given facility. This information can be queried as well, for example.

\vskip 0.1in\noindent
Key elements of the required infrastructure for a future query-based analysis system are expected to include:

\vskip 0.1in\noindent
\textbf{\textit{Sharing resources with traditional systems}}. Unlike a traditional batch system, access to this query system is intermittent, so it would be hard to justify allocating exclusive resources to it. Even with a large number of users to smooth out the minute-by-minute load, a query system would have a strong day-night load difference effect, weekday-weekend effect, and pre-conference effect. Therefore, the query system must share resources with a traditional batch system (performing event reconstruction for instance). Then the query system could elastically scale in response to load, preempting the batch system.

\vskip 0.1in\noindent
\textbf{\textit{Columnar Partitioning of Analysis Data}}.
Organizing data to enable fast-access of hierarchical event information (``columnar'' data) is both a challenge and an opportunity. Presenting column partitions to analysis systems as the fundamental unit of data management as opposed to files containing collections of events would bring several advantages for HEP end-user analysis (in contrast to reconstruction jobs). These column partitions would become first-class citizens in the same sense that files are today: either as single-column files or more likely as binary blobs in an object store. We note that columns are already a first-class citizen in the ROOT file system, however, appropriate data management and analysis software that leverages this capability is missing. Given a data store full of columns, datasets become loose associations among these columns, with metadata identifying a set of columns as mutually consistent and meaningful for analysis.

\vskip 0.1in\noindent
\textbf{\textit{Fast Columnar Data Caching}}.
Columnar cache is a key feature of the query system, retaining input data between queries, which are usually repeated with small modifications (intentionally as part of a systematics study or unplanned as part of normal data exploration). RAM cache would be a logical choice, given the speed of RAM memory, but the query system can't hold onto a large block of RAM if it is to share resources with a batch system. Furthermore, it can't even allocate large blocks of RAM temporarily, since this would trigger virtual memory swapping to a disk that is slower than the network it is getting the source data from. The query system must therefore stay within a tight RAM budget at all times. The query system's cache would therefore need to be implemented in SSD (or some future fast storage, such as X-Point). We can assume the query system would have exclusive access to an attached SSD disk, since caching is not required for the batch process.

\vskip 0.1in\noindent
\textbf{\textit{Provenance}}. The query system should also attach enough provenance to each dataset that it could be recreated from the original source data, which is considered immutable. User datasets, while they can't be modified in-place, can be deleted, so a dataset's paper trail must extend all the way back to source data. This paper trail would take the form of the original dataset name followed by queries for each step of derivation: code and closure data.

\vskip 0.1in\noindent
Additional details regarding the proposed query-based analysis approach, requirements and columnar management for HEP data can be found in the contributed CWP white paper ``Hierarchical Data Rolled Up in Columnar Arrays''~\cite{Pivarski2017}. 

\section{Analysis Preservation}

Reproducibility is the cornerstone of scientific results and while HEP does not face a reproducibility crisis, it is currently difficult to repeat most HEP analyses after they have been completed. This difficulty mainly arises due to the number of scientists involved, the number of steps in a typical HEP analysis workflow, and the complex ecosystem of software that HEP analyses are based on. In addition to the over-arching desire for scientific reproducibility, the ability to preserve and reproduce analyses also has immediate practical benefits to the collaborations and the LHC program. In particular, collaborations often want to extend completed analyses with more data, but this typically involves a lot of turnover in the analysis team. Similarly, if something anomalous shows up in the data, we often want to go back to an earlier version of the analysis and scrutinize it. Traditionally, both of these tasks have been difficult.  

Analysis preservation and reproducibility strategies are described in the Data and Software Preservation White Paper~\cite{HSF-CWP-2017-06}. Recent progress using workflow systems and containerization technology have rapidly transformed our ability to provide robust solutions. Nevertheless, analysis capture is best performed while the analysis is being developed in the first place. Thus, reproducibility needs to be considered in all new approaches under investigation and needs to be a fundamental component of the system as a whole. These considerations become even more critical as we explore analysis models with more heterogeneous hardware and analysis techniques.

\section{Analysis Interpretation}

The LHC collaborations are pursuing a vast number of searches for new physics. Interpretation of these analyses sits at the heart of the LHC physics priorities, and aligns with using the Higgs as a tool for discovery, identify the new physics of dark matter, and explore the unknown of new particles, interactions, and physical principles. The collaborations typically interpret these results in the context of specific models for new physics searches and sometimes reinterpret those same searches in the context of alternative theories. However, understanding the full implications of these searches requires the interpretation of the experimental results in the context of many more theoretical models than are currently explored by the experiments. This is a very active field, with close theory-experiment interaction and with several public tools in development.

A Forum on the interpretation of the LHC results for Beyond Standard Model (BSM) studies~\cite{LHCInterpretationForum} was thus initiated to discuss topics related to the BSM reinterpretation of LHC data, including the development of the necessary public recasting tools\footnote{``Recasting'' is a term used for this style of reinterpretation~\cite{Cranmer:1299950}.} and related infrastructure, and to provide a platform for a continued interaction between the theorists and the experiments. 

The infrastructure needed for analysis reinterpretation is a focal point of other cyberinfrastructure components including the INSPIRE literature database~\cite{INSPIRE}, the HEPDatadata repository~\cite{HEPDataRepo} , the CERN Analysis Preservation framework~\cite{CAP}, and the REANA cloud-based workflow execution system~\cite{REANA}. Critically, this cyberinfrastructure sits at the interface between the theoretical community and various experimental collaborations. As a result, this type of infrastructure is not funded through the experiments and tends to fall through the cracks. Thus, it is the perfect topic for a community-wide, cross-collaboration effort.

One specific piece of infrastructure that is currently missing is an analysis database able to represent the many-to-many mapping between publications, logical labels for the event selection defining signal and control regions, data products associated to the application of those event selections to specific datasets, the theoretical models associated to simulated datasets, the multiple implementations of those analyses from the experiments and theoretical community created for the purpose of analysis interpretation, and the results of those interpretations.

The protocol for (re)interpretation is clear and narrowly scoped, which makes it possible to offer it as a service. This type of activity lends itself to the Science Gateway\footnote{Science gateways allow science and engineering communities to access shared data, software, computing services, instruments, educational materials, and other resources specific to their disciplines~\cite{SciGateway}.} concept, which we refer to as the Interpretation Gateway. Such reinterpretation services have been foreseen for several years, and now most of the necessary infrastructure is in place to create it\footnote{RECAST: a reinterpretation service~\cite{Cranmer:1299950}, DPHEP Status Report~\cite{Akopov:2012bm}, ATLAS Data Access policy~\cite{ATL-CB-PUB-2015-001}.}. Such an interpretation service would greatly enhance the physics impact of the LHC and also enhance the legacy of the LHC well into the future.

\section{Analysis Roadmap}

\subsection{1-year Time Frame}

\begin{itemize}
\item Enable new open source tools to be plugged in dynamically in the existing ecosystem and mechanisms to dynamically exchange parts of the ecosystem with new components.
\item Develop requirements and design a next generation analysis facility concept, incorporating fast caching technologies to explore a query-based analysis approach and open-source cluster management tools.
\item Finalize full support of python in our ecosystem including long term maintenance.
\item Evolve policies to minimize this effort by retiring less used components from the integration and validation efforts.
\item Establish a schema for the analysis database.
\item Interpretation Gateway: conceptualization integrating the analysis facility, analysis preservation infrastructure, data repositories, and recasting tools. 
\end{itemize}

\subsection{3-year Time Frame}

\begin{itemize}
\item Research a comprehensive set of {\it bridges} and {\it ferries}, where a bridge enables the ecosystem to use an open source analysis tool and a ferry allows to use data from the ecosystem in the tool and vice versa.
\item Analysis facility: conceptualization of an end-to-end low-latency response high-capacity analysis facility with an initial prototype.
\item Develop a prototype functional or declarative programming language model.
\item Evolve our policies on how to replace components with new tools, maybe external, and solicit the community help in bridging and integrating them.
\item R\&D in the area of non-event information handling.
\item Develop a functional prototype for the Interpretation Gateway.
\end{itemize}

\subsection{5-year Time Frame}

\begin{itemize}
\item Analysis facility: evaluate chosen architectures and verify design or provide input for corrective actions (data from the 3rd running period of the LHC should be used).
\begin{itemize}
\item In Year 5, a blueprint for remaining developments and system design should become available, in time for deployment.
\end{itemize}
\item Interpretation Gateway: evaluate design or provide input for corrective actions to enable an LHC legacy (re)interpretation gateway.
\end{itemize}

\sloppy
\raggedright
\clearpage
\printbibliography[title={References},heading=bibintoc]

\end{document}